\documentclass[iop]{emulateapj}

\usepackage{graphicx}
\newcommand{\etal}{et al.\ }
\newcommand{\eg}{e.g., }
\newcommand{\ie}{i.e., }

\shorttitle{Bayesian Ages of Individual White Dwarfs}
\shortauthors{O'Malley, von Hippel \& van Dyk}

\begin{document}

\title{A Bayesian Approach to Deriving Ages of Individual Field White Dwarfs}

\author{Erin M. O'Malley}
\affil{Siena College and Dartmouth College}
\email{Erin.M.O'Malley@dartmouth.edu}
\author{Ted von Hippel}
\affil{Embry-Riddle Aeronautical University}
\and
\author{David A. van Dyk}
\affil{Imperial College London}

\begin{abstract}

We apply a self-consistent and robust Bayesian statistical approach to
determining the ages, distances, and ZAMS masses of 28 field DA white
dwarfs with ages of approximately 4 to 8 Gyrs.  Our technique requires only
quality optical and near-IR photometry to derive ages with $<$ 15\%
uncertainties, generally with little sensitivity to our choice of modern
initial-final mass relation.  We find that age, distance, and ZAMS mass are
correlated in a manner that is too complex to be captured by traditional
error propagation techniques.  We further find that the posterior
distributions of age are often asymmetric, indicating that the standard
approach to deriving WD ages can yield misleading results.

\end{abstract}


\section{Introduction}

Age is one of the most fundamental of all stellar properties, yet it is far
more difficult to determine age precisely than the other fundamental
properties such as stellar mass, surface temperature, and luminosity.  In
part, this is because we determine age via indirect means that require us
to first collect more directly observable quantities.  But there is another
major impediment; we can usually only measure age for an aggregate or
system of stars.  For example, the most common method of measuring stellar
ages---fitting stellar isochrones to cluster photometry---requires that we
obtain photometry for hundreds or thousands of stars in order to derive a
single age for the entire system.  We generally cannot precisely fit
isochrones to single stars unless these stars are in rare stages of their
evolution, such as just leaving the main sequence.  Even in this case, we
require exquisite distances and probably independently-determined stellar
masses (\eg from double-lined eclipsing binaries, Grundahl \etal 2008) to
derive a precise age.  The other major technique for measuring stellar
ages---fitting white dwarf cooling models to the white dwarf luminosity
function (WDLF)---also requires groups of stars.  This technique has been
used to derive an upper age limit to the Galactic disk (Winget \etal 1987;
Oswalt \etal 1996; Leggett, Ruiz, \& Bergeron 1998; Knox, Hawkins, \&
Hambly 1999) and to derive the ages of open and globular clusters (Claver
1995; von Hippel, Gilmore, \& Jones 1995; Richer \etal 1998; von Hippel \&
Gilmore 2000; Kalirai \etal 2001; Andreuzzi \etal 2002; Hansen \etal 2004;
2007; Garcia-Berro \etal 2010).  Yet, if we could reliably determine ages
for individual field stars, we could more readily determine the onset of
star formation in each of the Galactic stellar populations.

In this paper, we focus on determining ages for individual WDs.  The
technique for deriving individual WD ages was advanced by Bergeron,
Leggett, \& Ruiz (2001), who plotted WD mass versus $T_{\rm eff}$ for WDs
and compared them to WD cooling models to derive individual stellar ages.
Their version of this technique requires accurate WD masses ($\sigma_M \le$
0.1 M$_{\sun}$) and temperatures ($\sigma_T \approx 150$ K) from
spectroscopy for the warm WDs with Balmer lines or from precise parallaxes
($\sigma_{\pi} \le$ 10\%) for the cool WDs.  Bergeron \etal were able to
derive individual stellar ages precise to $\sim$1 Gy for WDs with mass $>$
0.6 M$_{\sun}$.  For WDs with mass $<$ 0.6 M$_{\sun}$, WD ages are
degenerate.  This now-standard technique relies on measuring $T_{\rm eff}$
from photometry or spectroscopy and log(\emph{g}) from spectroscopy or WD
surface area from trigonometric parallax.  Because WDs have a mass-radius
relation (Hamada \& Salpeter 1961), either log(\emph{g}) or surface area
yield mass, and the mass and $T_{\rm eff}$, when compared to a WD cooling
model, yield the WD cooling age.  The WD mass is relied upon again to infer
its precursor mass through the imprecisely-known initial-final mass
relation (IFMR), which is the mapping from masses on the zero age main
sequence (ZAMS) to WD masses.  The precursor mass is then converted to a
pre-WD lifetime via stellar evolution models (see Salaris \etal 2009, for a
list of models often used).  Finally, the precursor lifetime is added to
the WD cooling age to determine the total age of the WD.

The technique we have just outlined has its advantages and disadvantages.
The foremost advantage is that it yields reasonably precise ages for
individual WDs.  Secondarily, for cool WDs with masses $\geq$ 0.7
M$_{\sun}$, the progenitor lifetime is short relative to the WD cooling
age, and therefore uncertainties in the IFMR are unimportant (see, for
instance, von Hippel \etal 2006, fig 16).  On the negative side, this age
technique involves many steps, some of which are typically performed
inconsistently.  For example, Salaris \etal (2009) detailed how current
IFMRs may be constructed from an inconsistent set of isochrones or the
subsequent analysis may not use the same isochrones as those used in the
IFMR.  The IFMR itself is determined via WDs in star clusters.  Researchers
can reliably deduce the WD masses from spectroscopic log(g) measurements
(\eg Salaris \etal 2009; Williams, Bolte, \& Koester 2009), but determining
their precursor masses is model-dependent.  That model-dependent path
requires researchers to fit the main sequence turn-off with model
isochrones to derive the cluster age, subtract the WD cooling ages from the
total cluster age, derive precursor lifetimes, and then infer the precursor
masses from the same stellar evolution models.  Yet most IFMR studies
measure the WD masses and collect the cluster ages from the literature,
thereby using a heterogeneous mix of stellar evolution models (see
criticisms in Salaris \etal 2009) to infer precursor masses.  This
heterogeneous mix of stellar evolution models may or may not include the
model set that any subsequent researchers use to estimate a precursor
lifetime for their analysis of the age of an individual WD.  An additional
negative to determining WD ages via this process is that one has to
correctly propagate errors through many steps.  Some errors may start out
symmetrically distributed (\eg $T_{\rm eff}$), but we will show that the
assumptions behind the standard propagation of errors are not met, casting
doubt on the estimates and errors that they produce.

Our goal in this paper is to improve upon the current step-wise and often
internally inconsistent approach to obtaining individual WD ages.  We
accomplish this by applying the first model-based statistical analysis that
simultaneously fits WD photometry to models that combine stellar precursor
evolution, the IFMR, WD cooling, and WD atmospheres.

Our Bayesian statistical approach allows us to combine external
information, \eg distances from trigonometric parallaxes or spectroscopic
metallicities, with stellar photometry in order to calculate not only
reliable fitted values of stellar parameters, but also the entire posterior
distribution for each parameter, including error bars and correlations
among parameters.

\section{Observations of Hydrogen Atmosphere Field WDs}

Our technique can be applied to any WD for which we know the atmosphere
type and for which we have reliable models.  As a first test of our
technique, we sought a homogeneous sample of old H-atmosphere (DA) WDs with
optical and near-IR photometry.  DAs are the most common (Kleinman \etal
2004) and well-studied WDs, so they were a good starting point.  The 28 DAs
with optical and near-IR photometry published by Kilic \etal (2010) fit our
needs.  Kilic \etal selected 130 WDs from the large sample of Harris \etal
(2006) by targeting all WDs with bolometric magnitudes greater than 14.6
and tangential velocities greater than 20 km s$^{-1}$, the goal of which
was to create a clean sample of intrinsically faint and therefore old WDs.
Kilic \etal (2010) measured $JHK$ photometry for 126 stars in this SDSS
sample using the Near Infra-Red Imager and Spectrometer on Gemini-North,
the 0.8-5.4 micron medium-resolution spectrograph and imager on the
Infrared Telescope Facility, and the Wide-Field Camera on the United
Kingdom Infra-Red Telescope.  Their typical photometric error was 0.04 mag.
Although SDSS $u$ photometry is also available for these stars, because the
WD models we use (see below) did not fully incorporate the red wing of the
Lyman $\alpha$ line (Kowalski \& Saumon 2006; Rohrmann, Althaus, \& Kepler
2011), we chose not to incorporate these $u$-band data in our analyses.

Some of the cooler WDs in the Kilic \etal (2010) sample are undoubtedly
DAs, but they are too cool to excite Balmer lines, so their spectral type
is currently unknown.  Because our goal is to present our Bayesian
technique and outline its capabilities, we put off to a subsequent paper
the analysis of He-atmosphere (DB) WDs and WDs of uncertain spectral type.
The data we analyze therefore consists of $grizJHK$ photometry from Kilic
\etal for 28 DA WDs.

\section{Statistical Method}

We have developed a Bayesian approach to fitting isochrones to stellar
photometry (von Hippel \etal 2006; DeGennaro \etal 2009; van Dyk \etal
2009; Stein \etal 2013).  We term our software package BASE-9 for {\bf
B}ayesian {\bf A}nalysis of {\bf S}tellar {\bf E}volution with {\bf 9}
Parameters.  BASE-9 compares stellar evolution models (listed below) to
photometry in any combination of photometric bands for which there are data
and models.  BASE-9 was designed to analyze star clusters and accounts for
individual errors for every data point, ancillary data such as cluster
membership probabilities from proper motions or radial velocities, cluster
distance (\eg from Hipparcos parallaxes or the moving cluster method),
cluster metallicity from spectroscopic studies, and it can incorporate
information such as individual stellar mass estimates from dynamical
studies of binaries or spectroscopic atmospheric analyses of WDs.  BASE-9
uses a computational technique known as Markov chain Monte Carlo to derive
the Bayesian joint posterior probability distribution for six parameter
categories (cluster age, metallicity, helium content, distance, and
reddening, and optionally a parametrized IFMR) and brute-force numerical
integration for three parameter categories (stellar ZAMS mass, binarity, and
cluster membership).  The last three of these parameter categories include
one parameter per star whereas the first six parameter categories refer to
the entire cluster.  As a result, for star clusters BASE-9 actually fits
hundreds or thousands of parameters (= 3 N$_{\rm star}$ + 6)
simultaneously.  While we cannot constructively apply BASE-9 to individual
main sequence stars, we can profitably apply BASE-9 to individual WDs
because WDs have a mass-radius relation that constrains WD luminosity.  In
many cases, this constraint is sufficient to yield useful WD ages.  We
provide further details on our statistical method and computational
techniques in the appendix.

We used BASE-9 to fit model WD spectral energy distributions (SEDs) to the
observed $grizJHK$ photometry for 28 DAs.  BASE-9 generates model SEDs by
combining the following ingredients: stellar evolution models for the main
sequence through the asymptotic giant branch stage (Girardi \etal 2000; Yi
\etal 2001; or Dotter \etal 2008), an IFMR (Weidemann 2000; Williams \etal
2009; or one of two from Salaris \etal 2009), WD interior cooling models
(Montgomery \etal 1999; Renedo \etal 2010), and WD atmosphere models
(Bergeron \etal 1995).  The precursor stellar evolution models affect our
estimate of the length of time that a star spends evolving prior to
becoming a WD through the well-known strong dependence on mass and weak
dependence on stellar abundance.  Each of these three stellar evolution
models show only minor differences in precursor ages over the parameter
ranges we explore.

Because our statistical model is Bayesian, we can take advantage of prior
information, where available, to constrain parameters.  For this problem,
the results are insensitive to the precise choice of reasonable priors;
details are given in the appendix.  For stellar abundances, we assume that
all of these WDs are Galactic disk stars that started out with solar-ratio
abundances and a Gaussian distribution for metallicity, [Fe/H] = 0.0 $\pm$
0.3.  This is reasonable because these stars display disk proper motions
(Kilic \etal 2010).  Additionally, stellar abundances change the properties
of WDs only through slight changes to the precursor lifetimes (see
discussion below).  Because of the high surface gravities of WDs, their
primordial abundances are not reflected in their atmospheres (Dupuis \etal
1992; Koester \& Wilken 2006) and those WDs that are metal-polluted
(spectral type DZ, DAZ, or DBZ) are actively accreting from their
circumstellar environment (\eg Jura 2003; von Hippel \etal 2007; Farihi
\etal 2010).  We do not include DZ-type WDs in this analysis.  We set a
Gaussian prior on distance modulus, $m-M$ = 4.0 $\pm$ 2.5, equivalent to $63\pm^{136}_{43}$ pc.  Our distance prior is so loose that it does not constrain
the results.  It is included only because a prior on every parameter is
required by the Bayesian approach.  We impose a strict prior on absorption,
$A_{\rm V}$ = 0.  This absorption prior may be in error by a few
thousandths of a mag, but all of these stars are nearby, typically $<$ 100
pc, and out of the Galactic plane, so the absorption is essentially zero.

Besides our primary goal of developing a simultaneously consistent and
statistically robust method to derive ages for individual WDs, we set a
secondary goal of checking the sensitivity of WD ages to the IFMR.  This
sensitivity has been largely unexplored, yet has remained a caveat in many
studies involving WD ages (see Salaris \etal 2009).  We have already
studied the sensitivity of WD ages to stellar evolution models (DeGennaro
\etal 2009).  Therefore, in order to simplify things, rather than analyzing each of the 28 DAs with each of three stellar evolution models, each of four IFMRs, and both WD
cooling models, we chose a single stellar evolution model.  This allows us to work with eight results per WD, rather than two dozen.

The stellar evolution models we chose were from the Dartmouth Stellar Evolution
Database (DSED; Dotter \etal 2008).  The DSED models span a wide range of
parameter space including a metallicity range of $-2.5 <$ [Fe/H] $< +0.5$
and ZAMS masses from 0.1 to 4.0 M$_{\odot}$.  Because the upper mass limit
for WD precursors using the IFMRs we employ extend to 8.0 M$_{\odot}$ and
some of our WDs appear to have precursors more massive than 4 M$_{\odot}$,
we extrapolate precursor lifetimes for higher mass stars.  This
extrapolation introduces minimal error because the progenitor lifetimes of
these massive stars are so short.  A 5 and an 8 M$_{\odot}$ star, for
example, evolve to the WD stage in $\sim$120 and $<$ 60 Myrs, respectively
(Girardi \etal 2000).  Assuming our extrapolation technique were off by an
overly-conservative 50\% of the actual value as derived by stellar models
that went to higher masses, this would represent an error of $\leq$ 60 Myrs
for our stars.  As we will see below, the 28 DAs we analyze have ages of $\sim$4
to 8 Gyrs, so this extrapolation should introduce an age error of
typically $\leq$ 1\%.  At this point, that error is too small to force our
analysis to an isochrone set with such young ages.  Any of the three
stellar evolution model sets were suitable for our purposes and we
arbitrarily chose the Dotter \etal models for this analysis.

We explore results based on both the Montgomery \etal (1999) and the Renedo
\etal (2010) WD cooling models.  The Montgomery \etal models span the mass
range 0.4 to 1.2 M$_{\odot}$ and cooling age range from 0.3 Myr to 5.3-13.7
Gyr, depending on WD mass.  For WD masses greater than 1.2 M$_{\odot}$, we
extrapolate the Montgomery \etal models and for masses greater than 1.1
M$_{\odot}$ we occasionally must extrapolate the WD cooling ages.  The
Renedo \etal models span the mass range 0.524 to 0.934 M$_{\odot}$ and
cooling age range from 0 Myr to 9.4-18.3 Gyr, depending on WD mass.  For WD
masses greater than 0.934 M$_{\odot}$ we extrapolate the Renedo \etal
models.  Both model sets include realistic initial carbon/oxygen
distributions, the release of latent heat from crystallization, and the
gravitational energy liberated during carbon-oxygen phase separation upon
crystallization.  The Renedo \etal models were calculated for slightly
sub-solar metallicity (Z=0.01) and include non-grey atmospheres as boundary
conditions.  The Montgomery \etal models employ grey atmospheres for their
boundary conditions, which may be a limiting factor for those WDs with
$T_{\rm eff} <$ 6000 K.  The slightly subsolar metallicity for the Renedo
\etal models should have little effect on the implied WD ages for the
reasons we outlined about regarding the insensitivity of WDs to their
precursor metallicities and because we calculate the precursor lifetimes
independently using the DSED models for metallicities essentially from
within the prior on this parameter.

BASE-9 currently includes the four IFMRs cited above, each of which we use
in our study.  Salaris \etal (2009) derive two different IFMRs based on
observations of WDs in open clusters.  Salaris \etal paid particular
attention to consistency issues and accounted for uncertainties in WD
masses, WD cooling times, and progenitor masses.  The analysis behind the
Weidemann (2000) IFMR is not as sophisticated as that used by Salaris et
al., but because this IFMR is widely used, we have incorporated it as well.
The IFMR of Williams \etal (2009) adds significantly to the high-mass end
of the empirical IFMR by incorporating WDs from the young open cluster M35.
It is likely that none of these IFMRs are definitive, and in fact we (Stein
\etal 2013) are working on our own Bayesian approach to this problem.
Nevertheless, because the above-mentioned IFMRs are widely used and because
they are likely to approximately span the space occupied by the actual
IFMR, we have chosen to study these four relations.

\section{Results}

Of the nine possible parameters that we could fit with BASE-9, two are
meaningless (ZAMS mass of secondary companion and whether or not the star
is a cluster member) and three are set ($A_{\rm V}$ = 0; Y = 0.245 + 1.6 Z,
which is a standard helium-to-heavy-element relationship built into the
DSED models; and the IFMR is set to one of the four above-mentioned IFMRs,
rather than fitting our own).  Thus, for all stars in our sample, BASE-9
fits four parameters: the total stellar age, the precursor mass as it was
on the ZAMS, the initial metallicity, and the distance.  Figure 1 displays
contours of the posterior distributions for three WDs (J0003$-$0111,
J2045+0037, J2147+1127) projected onto four of the six possible
two-dimensional parameter planes based on the Montgomery \etal WD models.
Figure 2 is identical to Figure 1 except that the calculations employ the
Renedo \etal WD models.  We selected these three WDs because they are
representative of the range of posterior distributions.  Because the data
are uninformative for metallicity, the posterior and prior distributions
are indistinguishable, and we present only one of these three planes to
demonstrate that the posterior metallicity distribution essentially follows
the prior we set (see also further discussion below).

For all 28 WDs, the observed SEDs tightly constrain $T_{\rm eff}$, yet
weakly constrain WD mass.  As we can see in Figures 1 and 2, this is often
sufficient to yield reasonably tight age distributions.  The posterior
distributions in the distance-ZAMS mass plane are essentially the WD
mass-radius relation folded through the Stefan-Boltzmann luminosity
dependence on radius around a constrained $T_{\rm eff}$.  The age-ZAMS mass
and age-distance posterior distributions are substantially more complicated
and typically show three distinct regions.  The region of lowest ZAMS mass
stretching to greatest age is the region of parameter space where small
changes in WD mass change the total WD age primarily through the precursor
lifetime.  The region at intermediate masses typically displays the
opposite age-mass slope and is where small changes in WD mass affect WD age
primarily through changing the heat capacity of the cooling WD.  The
highest mass portion of the posterior distribution is where small changes
in WD mass change the WD age primarily through changing the contribution of
carbon or oxygen crystallization.  The high mass WD J2147+1127 is the most
constrained in these posterior distributions.  This is consistent with the
analysis of Bergeron \etal (2001), though BASE-9 required neither
independent distances from trig parallaxes nor a WD mass determination from
spectroscopic fits to log($g$).  The Montgomery \etal and Renedo \etal fits
are broadly similar, though with differences in the detailed shapes of the
distributions, particularly for the two lower mass stars near ZAMS masses
of 4 M$_{\odot}$.  The mean ages also shift between the Montgomery \etal and
Renedo \etal fits.  We return to a comparison between these two model sets
later.  Taken together, these diagrams show strong asymmetric posterior
distributions, which is both a testament to the non-linearities of stellar
evolution and a warning of the potential pitfalls of standard error
propagation strategies.

\begin{figure}[h!]
\begin{centering}
\includegraphics[scale=0.115]{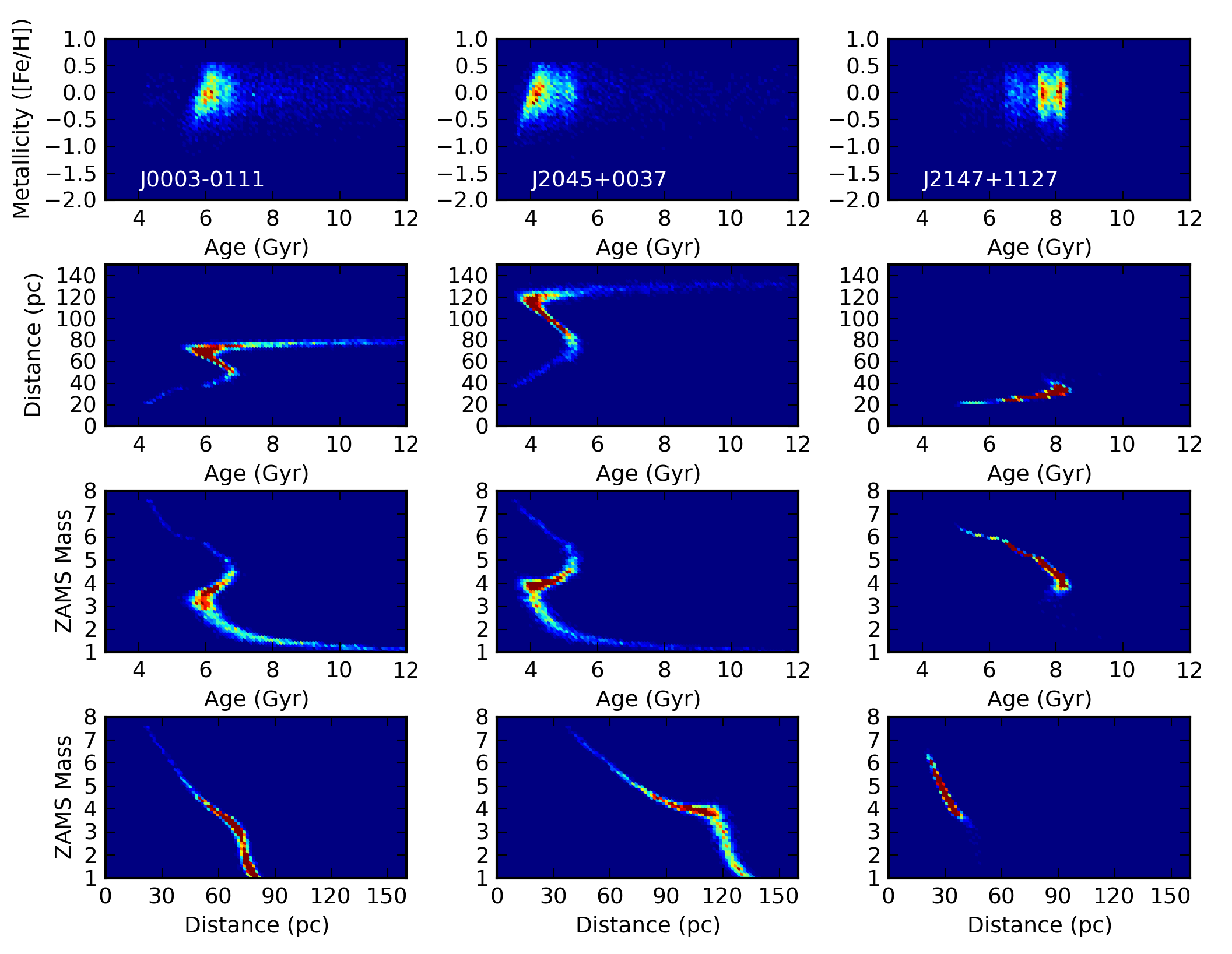}
\caption{Projections of the joint posterior distributions into (from top to
bottom) the age-metallicity, age-distance, age-ZAMS mass, and distance-ZAMS
mass planes for three representative WDs in our sample, all analyzed with
the Montgomery \etal WD cooling models and Williams \etal IFMR.  The stars
demonstrate posterior distributions for a WD with a typical age and mass
(J0003$-$0111, left-most column), a somewhat younger age and greater
distance (J2045+0037, middle column), and greater age and mass (J2147+1127,
right-most column).}
\end{centering}
\end{figure}

\begin{figure}[!h]
\includegraphics[scale=0.115]{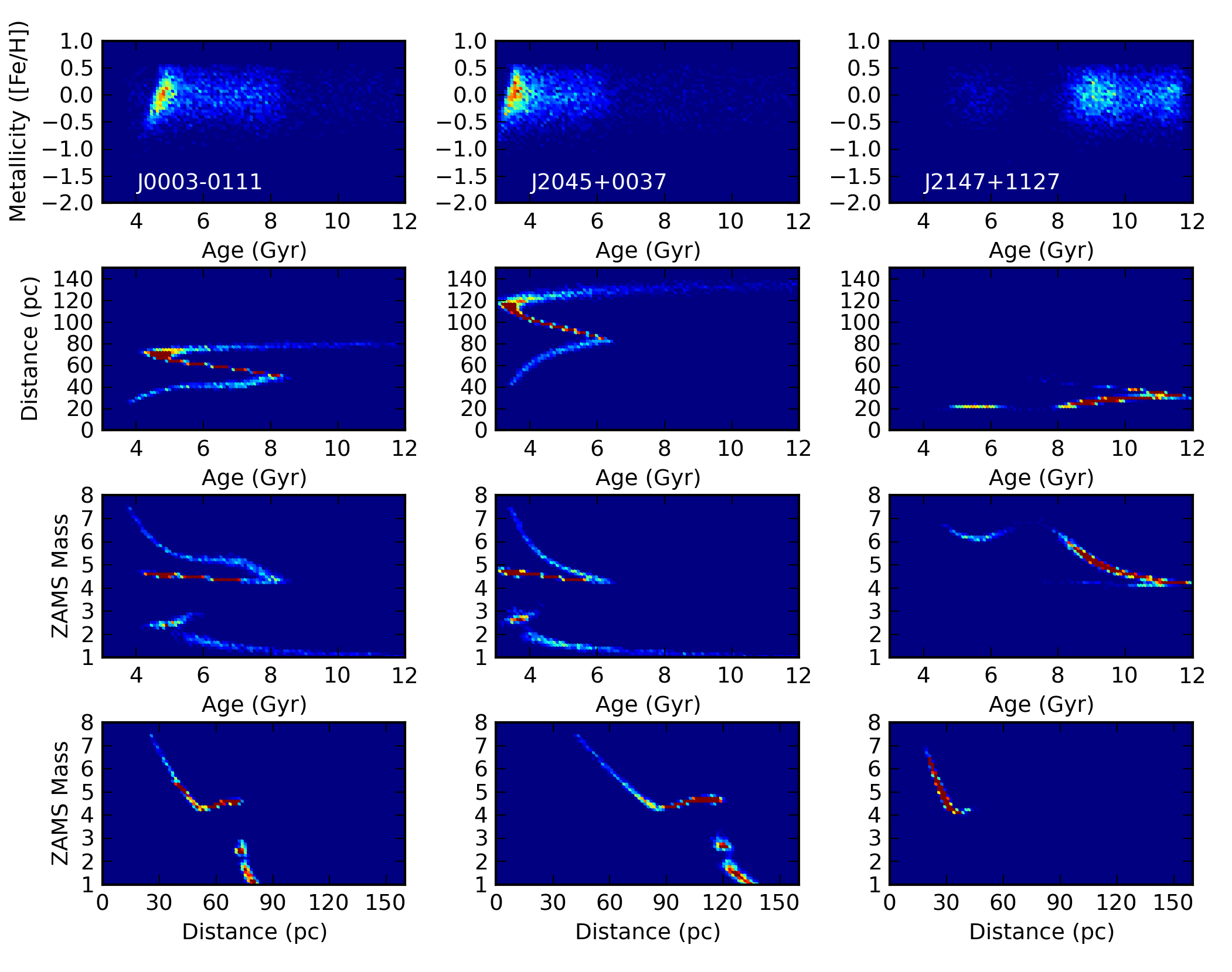}
\caption{Similar to Figure 1, except analyzed with the Renedo \etal WD
cooling models.  The gap in mass near 4 M$_{\odot}$ is a common feature 
of the Renedo \etal fits and is due to a steeper color vs.\ mass 
relationship in these models compared to the Montgomery \etal models
and a sudden change in this slope at a WD mass of 0.877 M$_{\odot}$,
corresponding to a ZAMS mass of $\sim$4 M$_{\odot}$, depending 
somewhat on the IFMR.}
\end{figure}

Figure 3 displays the posterior distributions for a single representative
star (J0003$-$0111), fit with the Montgomery \etal WD models and each of
the four IFMRs we have studied.  There are detailed similarities in all
four cases, and in fact the contours for all IFMRs peak near 6 Gyrs and 65
pc, yet the distributions are subtly different from one IFMR to another.
For instance, the upper distance limits extend $\sim$10 pc further for the
Weidemann IFMR and the Salaris \etal Piecewise IFMR than for the other
IFMRs and the lower distance extrema move even more substantially.  Such
comparisons for other stars also show broadly similar results from one IFMR
to another and in subsequent analyses we will compare summary statistics
for the different IFMRs, rather than the entire posterior distributions.

\begin{figure}[ht!]
\begin{centering}
\includegraphics[scale=0.115]{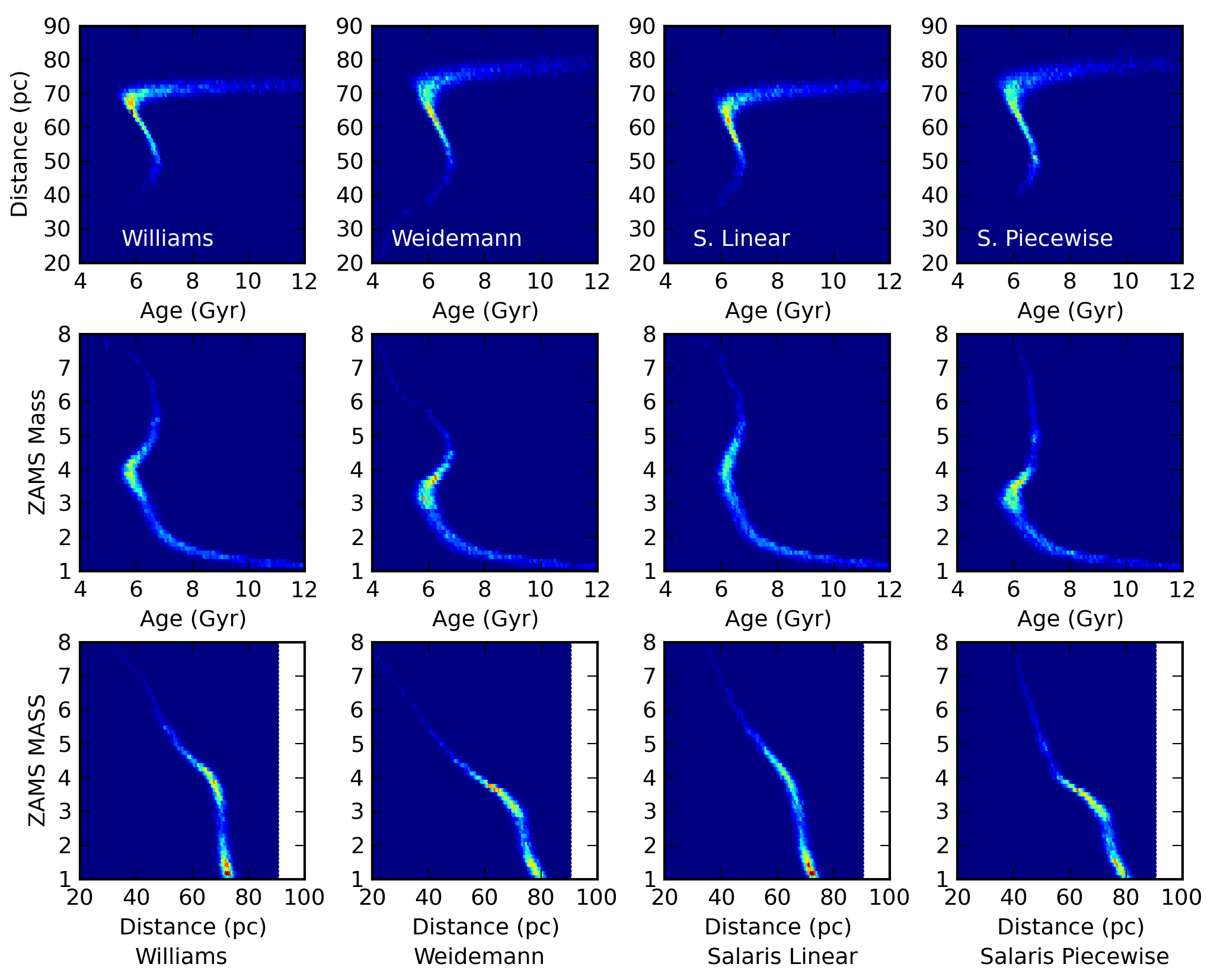}
\caption{Posterior probability projections in the age, distance, ZAMS mass
planes for J0003$-$0111 for each of four IFMRs and the Montgomery \etal WD
cooling models.  These IFMRs, from left to right, are the Williams et al.,
Weidemann, Salaris \etal Linear, and Salaris \etal Piecewise relations.}
\end{centering}
\end{figure}

Figure 4 displays the marginal (\ie collapsed, one dimensional) posterior
distributions for the four fitted parameters for the three WDs presented in
Figures 1 \& 2 for all four IFMRs.  The different IFMRs are color coded and
the solid and dashed histograms indicate fits based on Montgomery \etal and 
Renedo \etal WD models.  As expected, all of the metallicity distributions
for all stars using all IFMRs are the same and in fact are very close to
the priors on [Fe/H] (we find posterior values 0.27 $<$ $\sigma$([Fe/H])
$<$ 0.29).  The metallicity distributions are truncated at [Fe/H] = +0.5
because that is the upper metallicity limit of the Dotter \etal isochrones.
The age distributions are more complicated than the [Fe/H] distributions,
yet are broadly consistent among the IFMRs but not always between the two
WD models.  The distance and ZAMS mass distributions can be different
from one IFMR to another.  Because distance is a directly measurable
quantity, in principle coupling this type of analysis with precision trig
parallaxes for the right stars could rule for or against particular IFMRs
within some mass ranges.  For instance, for both J0003$-$0111 and
J2045$+$0037, greater distances are possible with some IFMRs than with
others.

\begin{figure*}[ht]
\begin{centering}
\includegraphics[width=4.8in]{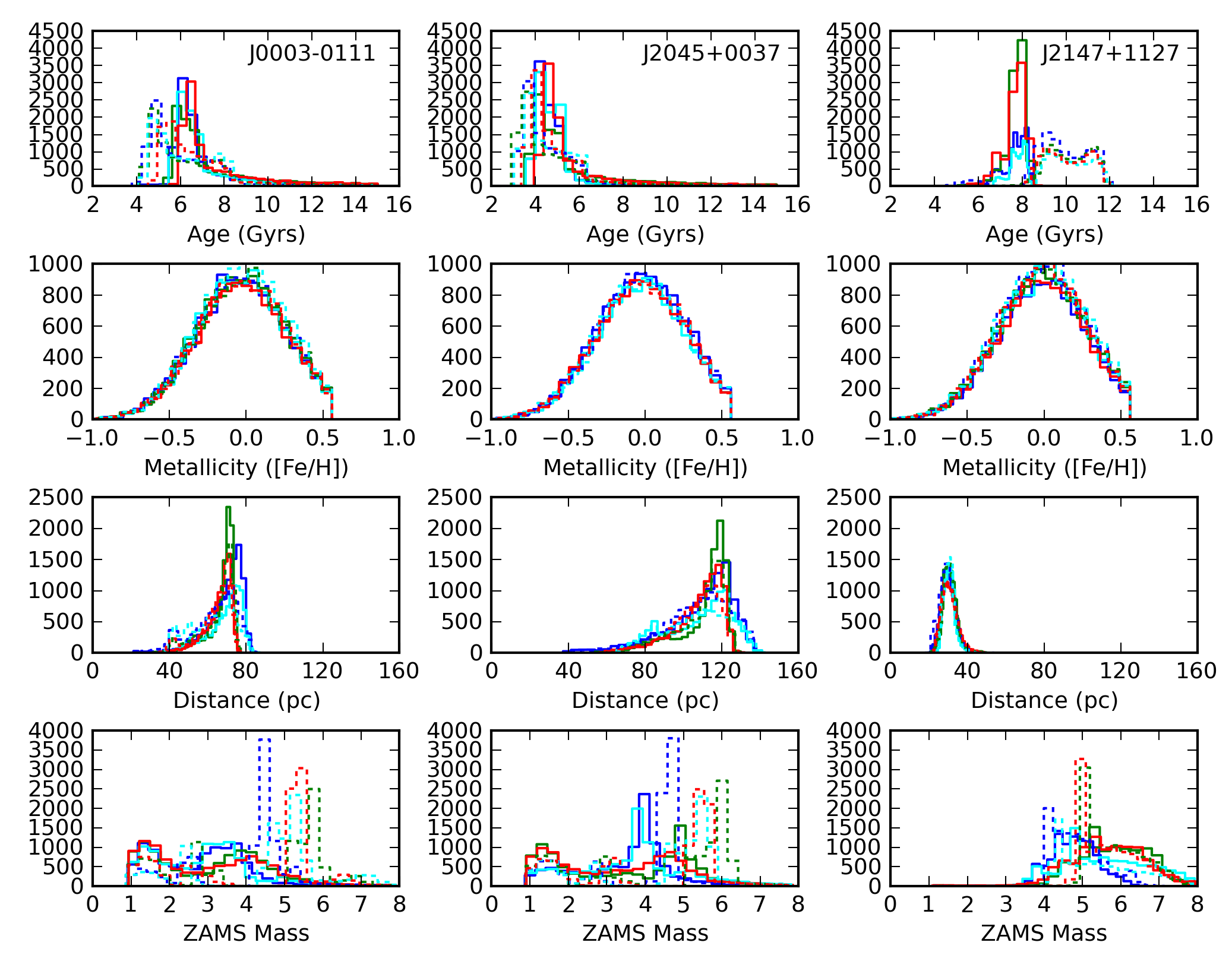}
\caption{Marginal posterior distributions for the four fitted parameters
for each of the three stars presented in Figure 1.  All four IFMRs are
presented as color-coded histograms with green for Weidemann, blue for
Williams et al., cyan for Salaris \etal Linear, and red for Salaris \etal
Peicewise Linear.  The Montgomery \etal models are indicated with solid
lines and the Renedo \etal results are indicated with dashed lines.}
\end{centering}
\end{figure*}

Figure 5 summarizes the age information for all 28 DAs that we have
analyzed using all four IFMRs and the Montgomery \etal WD models.  The
horizontal axes in all panels display age assuming the Williams \etal IFMR
and the vertical axes display the age difference under Williams \etal and
under the other IFMRs.  The points and error bars indicate both the average
and median ages along with the 68\% confidence intervals.  We note that
these confidence intervals are not derived to be symmetric about the median
or average, but rather mark the values beyond which the last 16\% of
distribution at each end resides.  Despite differences in the shapes of the
age distributions, for most of these old DA WDs, both the median and
average ages are essentially identical from one IFMR to another.  The
exception are 3 or 5 of the 8 youngest WDs, and to a lesser extent, 2 WDs
near 6.2 Gyr, all of which systematically differ in age between the
Williams \etal IFMR on the one hand and the other three IFMRs on the other
hand.  Overall this is good news given the current state of uncertainty in
the IFMR.  The average age uncertainty for these 28 DAs analyzed under the
Montgomery WD models is $\pm^{10.5\%}_{13.5\%}$.

\begin{figure}[h!]
\begin{centering}
\includegraphics[scale=0.108]{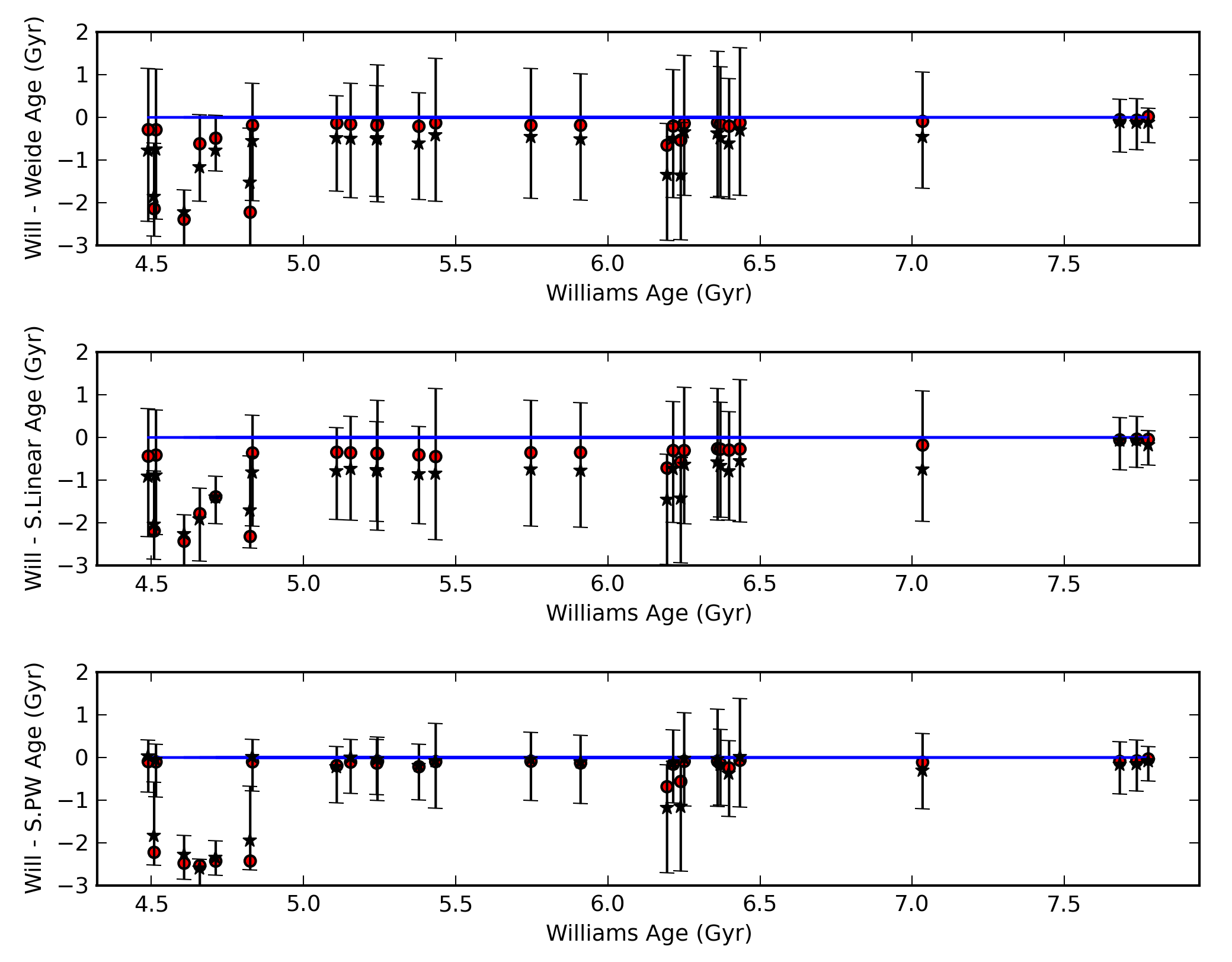}
\caption{Average (black star symbols) and median (red circle symbols)
differences in calculated ages for 28 DAs for each of four IFMRs for
the Montgomery \etal models.  The error bars represent the 68\% confidence
intervals in the marginalized posterior age distribution for the wider of
the two age fits.}
\end{centering}
\end{figure}

Figure 6 presents the differences in median distances for each of the 28
DAs as analyzed with each of the four IFMRs.  As with Figure 5, the error
bars represent the 68\% confidence interval of the distance posterior
distribution for the wider of the two distributions being compared.  Most
stars have statistically similar median and average distances no matter
which IFMR is used, but there are differences among some of the same stars
that were inconsistent in Figure 5.  This figure reiterates a point from
Figure 4, which is that for some stars, follow-up precision distances could
rule for or against any particular IFMR within a particular mass range.

\begin{figure}[h!]
\begin{centering}
\includegraphics[scale=0.108]{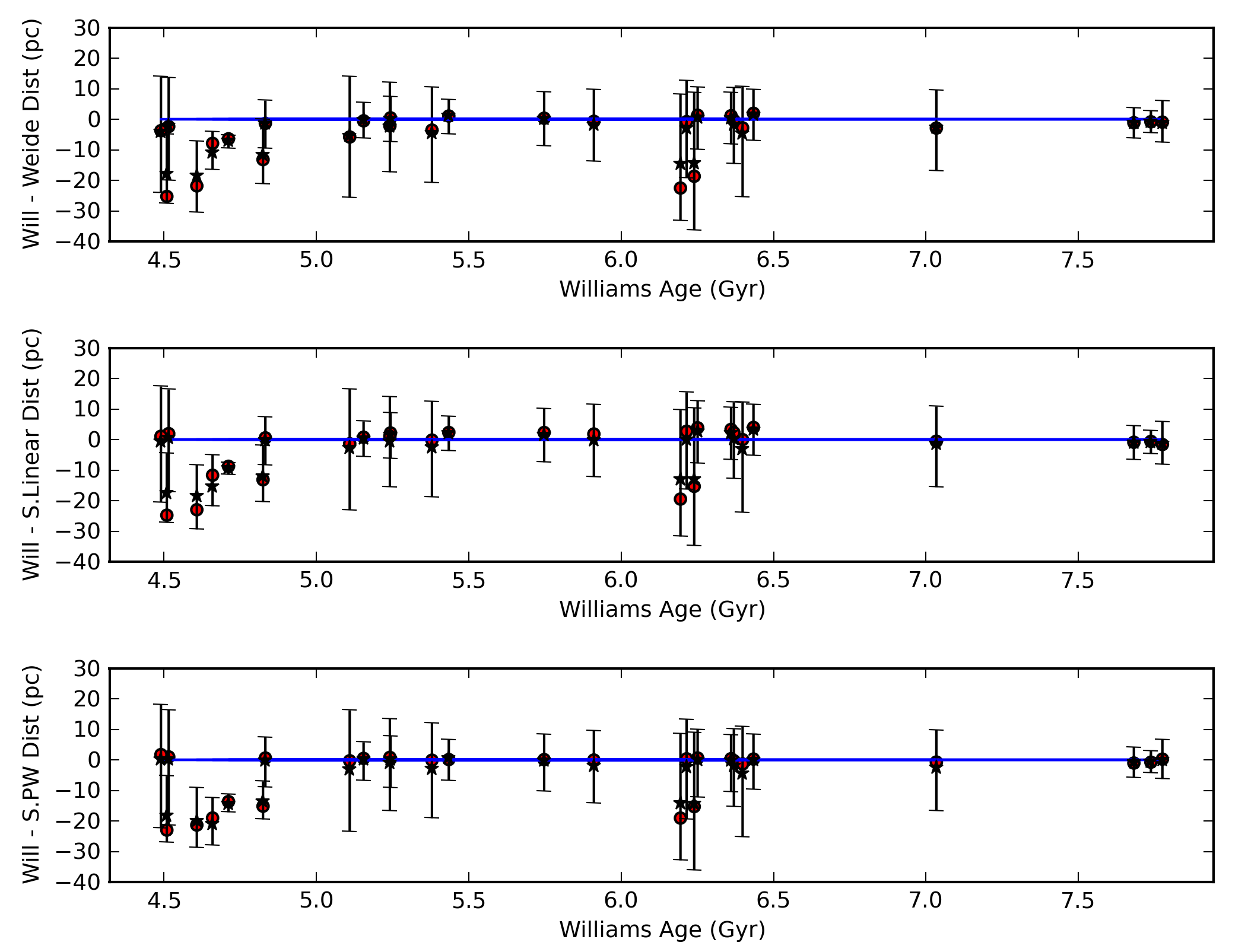}
\caption{Similar to Figure 5, but now the y-axis displays the difference
in distance fits for each of the four IFMRs.}
\end{centering}
\end{figure}

In Figure 7 we present the cumulative distributions of the fitted median WD
masses for our 28 DAs under each of the four IFMRs and both WD models.
These cumulative distributions are often different from each other,
particularly the Williams \etal IFMR versus the other IFMRs, and there is
an offset from one WD model to another.\footnote{The standard difference
statistic used in astronomy, the Kolmogorov-Smirnov test, is inappropriate
to check these differences, as this statistic is meant to check on
independent data presumably drawn from the same sample, rather than the
same data analyzed under different assumptions.}  In all cases, our sample
of 28 DAs likely contains a few high mass WDs.  The Williams \etal IFMR
fits imply that ten WDs may have masses greater than 0.9 M$_{\odot}$
(J074721+24, J0821+3727, J0947+4459, J1102+4030, J1130+1002, J1317+0621,
J1534+0711, J1722+2848, J2147+1127, and J2342$-$1001).  We remind the
reader that these masses are not directly fit and that a constraining prior
on distance could decrease any of these implied masses.  Yet, because of
the relative rarity of high mass WDs and the likely onset of
crystallization for stars of this age and implied mass (Metcalfe,
Montgomery, \& Kanaan 2004), these objects merit additional scrutiny.

\begin{figure}[t!]
\begin{centering}
\includegraphics[scale=0.108]{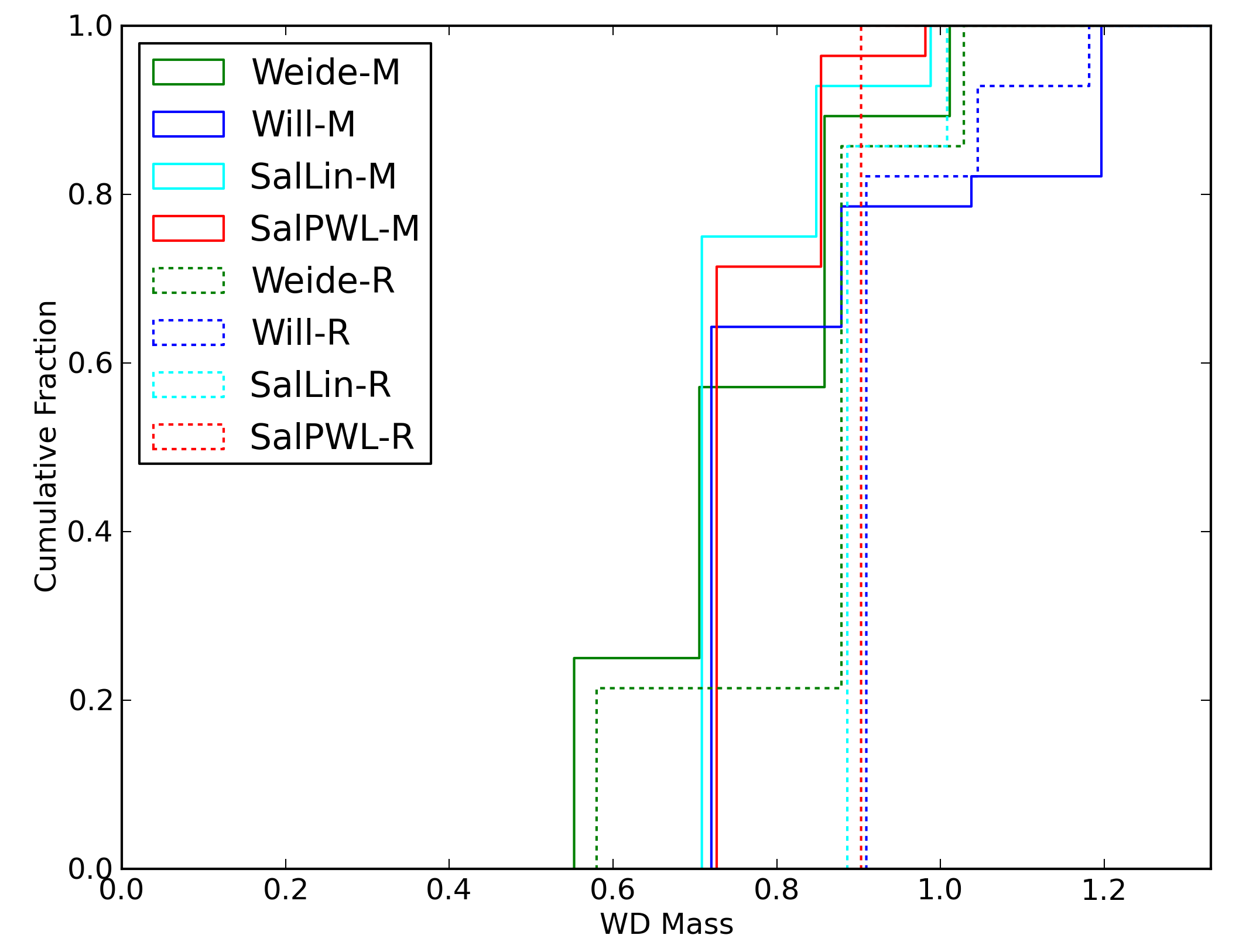}
\caption{Cumulative distribution of median WD masses for 28 DAs analyzed under
four IFMRs.  Line styles are the same as in Figure 4.}
\end{centering}
\end{figure}

We can make an initial quantitative comparison between the traditional
approach to deriving individual WD ages by comparing the ages derived by
Kilic \etal (2010) with our Bayesian results for these 28 WDs.  Kilic \etal
did not have access to spectroscopic log($g$) information, and thus did not
have masses for these stars.  Instead, they assumed log($g$) = 8.0, which
is a common approach in this situation and equivalent to assuming all WDs
have masses of $\sim$0.58 M$_{\odot}$ for this $T_{eff}$ range (Bergeron
\etal 1995).  Because WDs have a narrow mass peak near 0.6 M$_{\odot}$ (\eg
Liebert, Bergeron, \& Holberg 2005 who find a standard deviation around
this peak of $<0.2 M_\odot$), this approach should yield only a slight age
bias with some scatter introduced by the actual WD masses.  Additionally,
Kilic \etal report the WD cooling ages, whereas BASE-9 yields total WD
ages.  We thus calculated the posterior distribution of cooling ages for
each of these 28 WDs in order to compare our results and those of Kilic et
al. (2010).  Figure 8 displays that comparison with both the Montgomery
\etal (top panel) and Renedo \etal (bottom panel) WDs.  There is a
systematic offset of $\sim$ 1-2 Gyr for most stars broadly distributed
across age.

\begin{figure}[h!]
\begin{centering}
\includegraphics[scale=0.108]{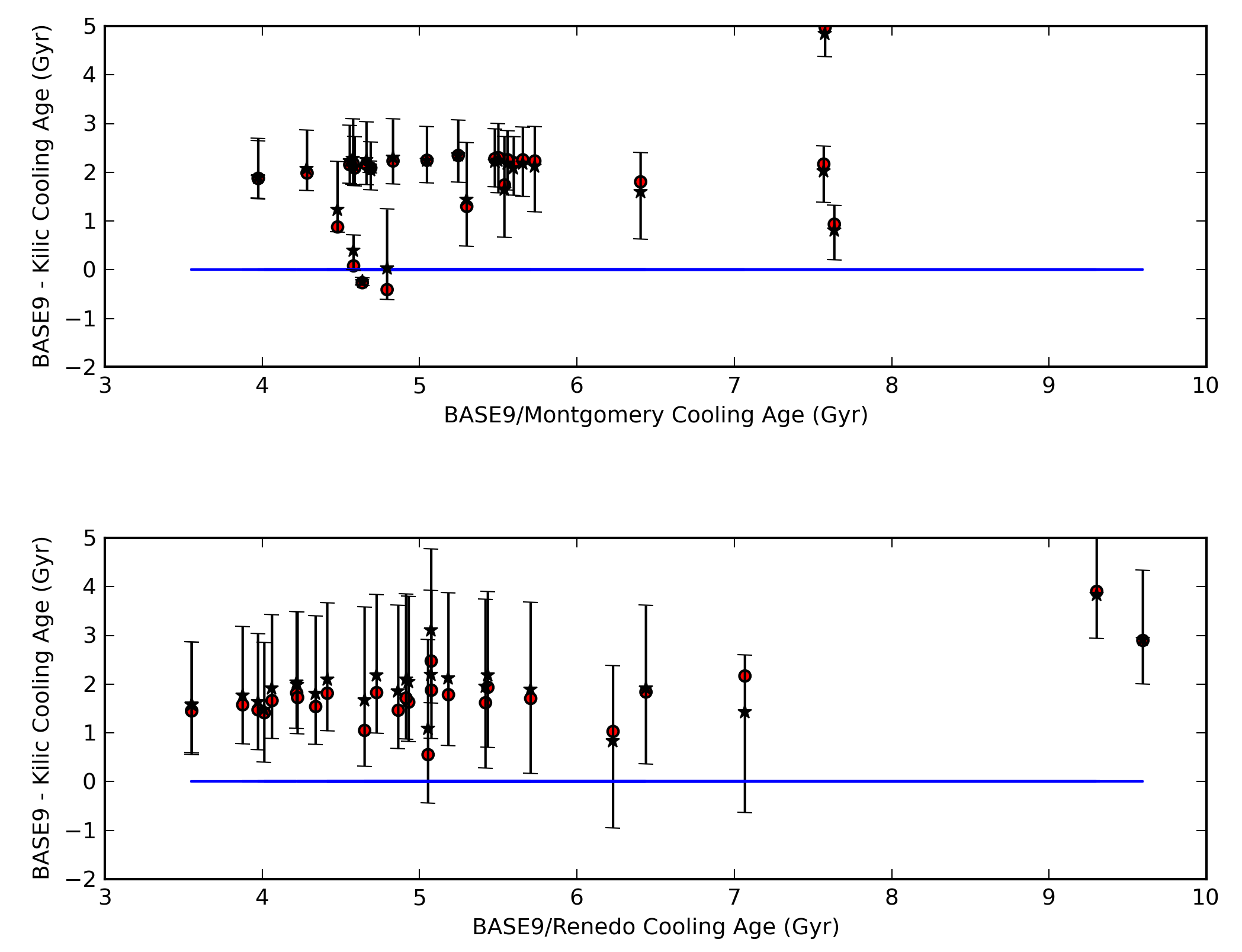}
\caption{Average (black star symbols) and median (red circle symbols)
differences between the BASE-9 cooling ages and the Kilic \etal cooling
ages.  The error bars represent the 68\% confidence intervals of the
BASE-9 cooling age posterior distribution.  Kilic \etal did not provide 
age uncertainties.}
\end{centering}
\end{figure}

If the entire systematic between our ages and the Kilic \etal ages were due
to the Kilic \etal assumption that log($g$) = 8, we should see a strong
correlation between the difference of our WD mass estimate and the Kilic
\etal mass assumption versus the difference in WD cooling ages (BASE-9
cooling age minus Kilic \etal cooling age).  We plot that comparison in
Figure 9.  There is no meaningful correlation in this diagram, so the
systematic age difference cannot be primarily due to the log($g$) = 8
assumption.  This bolsters the case that the traditional age determination
technique can give substantially different answers than our Bayesian
approach and thus that the standard, step-wise process for determining ages
of individual WDs may yield misleading results.

\begin{figure}[h!]
\begin{centering}
\includegraphics[scale=0.108]{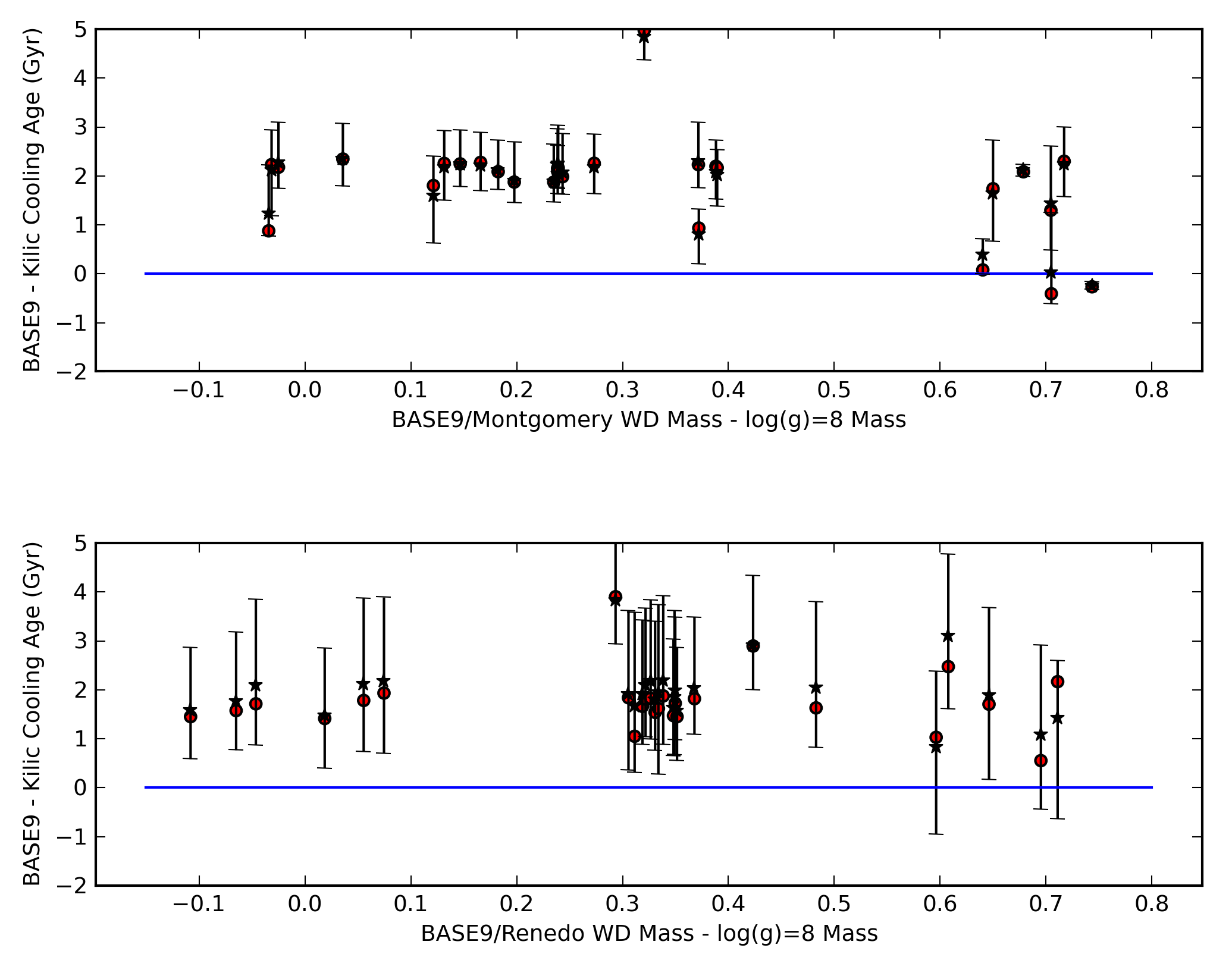}
\caption{The difference between the median BASE-9 WD mass estimates and the
log(g)=8 masses versus the difference in WD cooling ages between BASE-9 and
those derived by Kilic et al.  The symbols and error bars have the same
meaning as in Figure 8.}
\end{centering}
\end{figure}

Up until this point, we have noted a few differences among the results
based on whether we used Montgomery \etal or Renedo \etal WD models, but we
have not directly compared our results based on these two models.  In Figure 10
we provide this comparison employing the Williams \etal IFMR.  It is
comforting to see that these two modern WD cooling models yield consistent
results for 25 of these 28 WDs.  For the three WDs that are inconsistent,
they differ in the sense that the Renedo \etal models imply ages $\sim$2
Gyr older than the Montgomery \etal models.  Interestingly, these stars are
all cooler than $T_{\rm eff}$ = 5500 based on the Bergeron \etal model
atmosphere colors, and it is precisely in this region where theorists
expect non-grey atmospheres, as incorporated in the Renedo \etal models, to
be important in modelling cool WDs.  Unfortunately, we cannot ascribe this
difference to grey vs.\ non-grey atmospheres because 50+\% of the posterior
mass distributions for all three of these WDs lies beyond the Renedo \etal
upper mass limit (0.934 M$_{\odot}$), and therefore these fits with the
Renedo \etal models required extensive extrapolation in mass.

\begin{figure}[!t]
\begin{centering}
\includegraphics[scale=0.11]{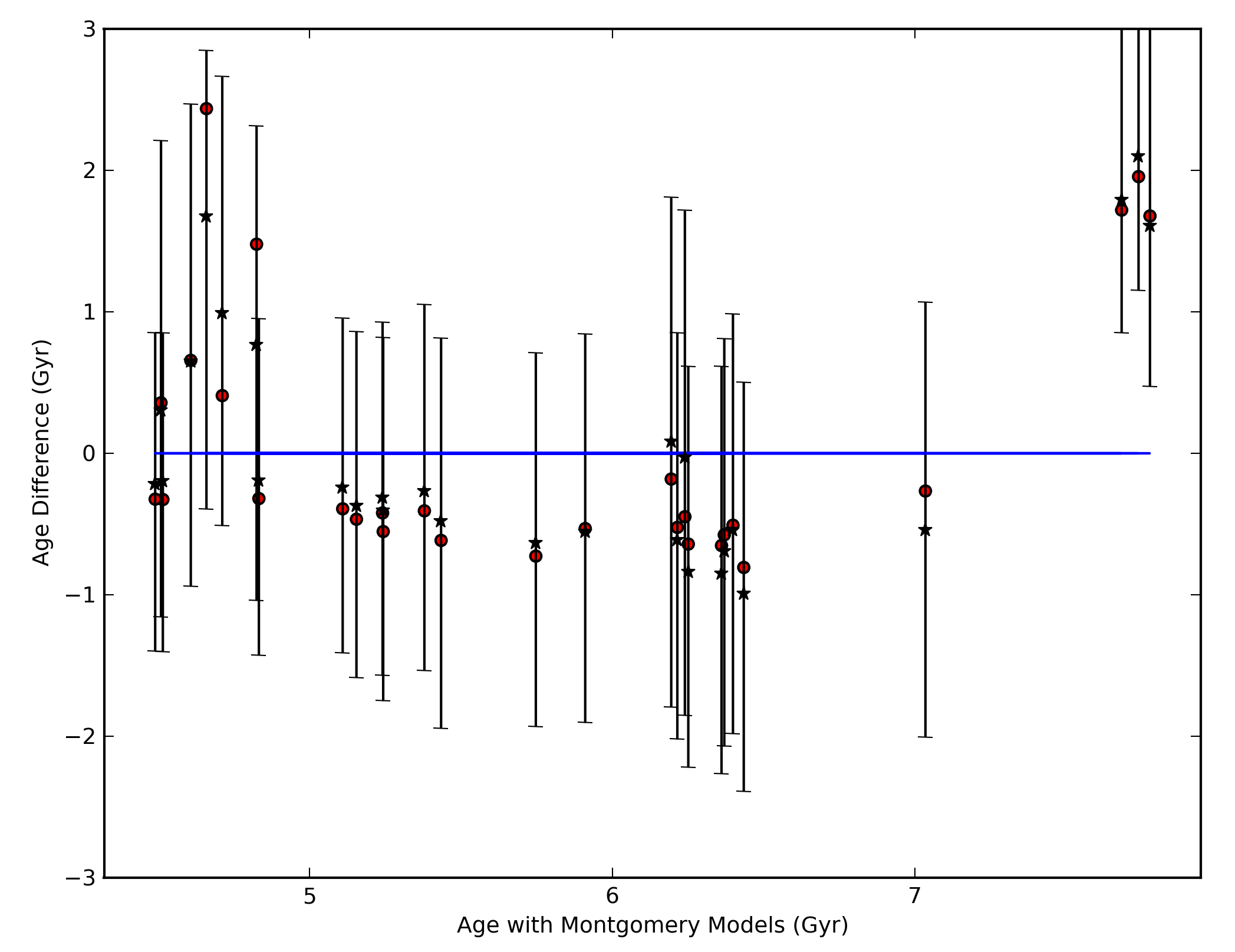}
\caption{Montgomery \etal ages versus Renedo \etal ages.  The vertical axis
gives the difference in the sense Renedo $-$ Montgomery, with the symbols
and errors bars as in Figure 5. These fits were made with the Williams
\etal IFMR.}
\end{centering}
\end{figure}

\section{Conclusions}

We have applied a self-consistent and robust Bayesian statistical approach
to determining the ages, distances, and ZAMS masses of individual WDs.  We
find that age, distance, and ZAMS mass are correlated in complicated
posterior distributions.  While these correlations make sense in terms of
the non-linearities of stellar evolution, they are too complex to be
quantified by traditional error propagation methods.  Additionally,
because the age posterior distributions are often asymmetric, traditional
techniques can yield misleading ages.

We find that for our DA sample in the age range of $\sim$4 to 8 Gyrs, that
SDSS $griz$ photometry supplemented by quality $JHK$ photometry is
sufficient to derive ages with errors $<$ 15\%.  Furthermore, these ages
are often (but not always) insensitive to which of the current modern IFMRs
one uses.  We expect that these age uncertainties could be substantially
reduced with additional information from spectroscopy or parallax
measurements that would constrain WD masses.

We also find that distances to some of the WDs in our sample could
rule for or against one or more of the IFMRs within particular mass ranges.
These distances could be incorporated into a new BASE-9 analysis to derive
a new principled estimate of the IFMR.  Trigonometric parallaxes for some
of these objects would be particularly valuable.

\acknowledgements

We thank Michael Montgomery for calculating WD sequences for use with BASE-9.  We thank Arthur Byrnes and the Embry-Riddle Aeronautical University High Performance Computing Cluster for supporting our calculations and Elliot Robinson for helping develop BASE-9.  We thank an anonymous referee and our editor, Eric Feigelson, for feedback that
substantially improved this paper.  This material is based upon work supported by the National Aeronautics and Space Administration under Grant NNX11AF34G issued through the Office of Space Science. In addition, David van Dyk was supported in part by a British Royal Society Wolfson Research Merit Award, by a European Commission Marie-Curie Career Integration Grant, and by the STFC (UK).\nopagebreak

\clearpage

\appendix{Statistical Method}

\newcommand{\bSigma}{\mbox{\boldmath{$\Sigma$}}}
\newcommand{\bTheta}{\mbox{\boldmath{$\Theta$}}}
\newcommand{\bX}{\mbox{\boldmath{$X$}}}
\newcommand{\bGWD}{\mbox{\boldmath{$G$}}_{\textrm{WD}}}

The statistical methods we use are a special case of those described in
Stein \etal (2013), see also DeGennaro \etal (2009) and van Dyk \etal
(2009).  Because our WDs are nearby and at high Galactic latitude, we fix
$A_{\rm V}=0$.  The helium abundance, important in the evolution of the WD
precursor, is set in the DSED models at $Y=0.245 + 1.6$Z.  We specify the
likelihood function for fitting the remaining stellar parameters,
$\bTheta=$ ($m-M$, [Fe/H], log(IMF), log(age)).  Specifically, for a single
white dwarf, this likelihood can be written,
$$
L(\bTheta | \bX, \bSigma)\nonumber = 
 {1\over \sqrt{(2\pi)^{n}|\bSigma|}}
 \exp\left(-{1\over2}\Big(\bX - \bGWD(\bTheta)\Big)^\top\bSigma^{-1}\Big(\bX - \bGWD(\bTheta)\Big)   \right) 
 \label{eq:like}
$$
where $\bX$ is the vector of observed photometric magnitudes, $\bSigma$ is
the variance-covariance matrix of the observation errors, and
$\bGWD(\bTheta)$ is the vector of predicted photometric magnitudes as a
function of the stellar parameters.  $\bGWD$ takes as input the mass of the
WD precursor and the total age of the star, passes that through the DSED
stellar evolution models, one of four IFMRs, one of two WD interior models,
and a WD atmosphere model to predict the photometric magnitudes.

Our Bayesian analysis is based on the posterior distribution of $\bTheta$,
namely
$$
p(\bTheta |  \bX, \bSigma) \propto L(\bTheta | \bX, \bSigma) p(\bTheta),
$$
where $p(\bTheta)$ is the prior distribution for the stellar parameters.
The prior distribution quantifies knowledge about the likely values of the
stellar parameters that we have before considering the current data set.
The posterior distribution combines this information with that in the data
and summarizes all of the available information including the current data.
We specify $p(\bTheta)$ independently for each of the stellar parameters;
details appear in Table~1.

Statistical inference for the stellar parameters is based on a Monte Carlo
simulation from their posterior distribution.  These simulations are
plotted to represent the posterior distribution, e.g., in Figures 1-3 or
summarized by their mean or median to estimate the parameters.  Although
the likelihood is based on a simple Gaussian distribution, the dependence
of its mean on the stellar parameters can be highly non-linear, potentially
leading to an irregular posterior distribution for $\bTheta$, see, e.g.,
Figures 1-3. We use a Markov chain Monte Carlo (MCMC) sampler on the joint
posterior distribution of $m-M$, [Fe/H], and log(age), which is obtained by
numerically integrating $p(\bTheta |\bX, \bSigma)$ over the log(IMF).  We
use a set of initial MCMC iterations to approximate the posterior
variance-covariance matrix of $m-M$, [Fe/H], and log(age) and then run a
Metropolis sampler that uses this matrix in its jumping rule.  We typically
thin the sample by a factor of 100-500 to obtain a subsample with lower
correlation. We only present results of WDs where the resulting MCMC
sampler, after appropriate burn in, appears to deliver a reliable
representation of the posterior distribution.  We recover the posterior
ZAMS mass distribution by sampling it from its conditional posterior
distribution after the MCMC run is complete.

We performed a sensitivity analysis on our priors by relaxing the two
($m-M$ and [Fe/H], see Table 1) for which we have no independent data.
Our nominal priors were relaxed to $m-M$ = 4 $\pm$ 5 ($63\pm^{568}_{57}$ pc) and
[Fe/H] = 0 $\pm$ 1.0.  We then reran 7 WDs with the Renedo \etal interior
models and the Williams \etal IFMR.  For these 7 WDs, although the average
metallicities decreased by $\sim$0.4 dex, because our models bound the
upper limit of [Fe/H] at $+$0.5, the other parameters differed very little.
The average ages differed by 0.040 to 0.167 Gyr, or 0.77\% to 3.17\%.  The
average distances differed by only 0.4 to 2.8 pc, and always $\leq$ 3.6\%.
The average ZAMS masses for these stars differed by 0.07 to 0.18
M$_{\odot}$, always $\leq$ 5\%.

 
\begin{table}[!h]
\begin{center}
\begin{tabular}[c]{l l}
\multicolumn{2}{p{8cm}}{\caption{Prior Distributions}} \\
\hline
Quantity & Prior \\
\hline
$m-M$    & Gaussian($\mu=4.0$, $\sigma=2.5$), equivalent to dist = $63\pm^{136}_{43}$ pc \\
Fe/H     & Gaussian($\mu=0.0$, $\sigma=0.3$) \\
IMF      & log(IMF) $\sim$ Gaussian($\mu=-1.02$, $\sigma=0.67729$) from Miller \& Scalo (1979) \\
         & subject to M$_{\rm WD,up}$, upper ZAMS mass limit to produce a WD = 8 M$_{\odot}$ \\
log(age) & Uniform above 250 Myrs, 0 below \\
\hline
\hline
\end{tabular}
\end{center}
\end{table}
\clearpage

\end{document}